\documentclass{kapproc} 
\let\footnote\savefootnote
\let\footnotetext\savefootnotetext 
\let\footnoterule\savefootnoterule 
\kluwerbib

\RequirePackage{graphicx}%
\RequirePackage{epsf}%

\newcommand\chandra{{\sl Chandra}}
\newcommand\xmm{{\sl XMM}}
\newcommand\rosat{{\sl ROSAT}}
\newcommand\asca{{\sl ASCA}}
\newcommand\sax{{\sl SAX}}
\newcommand\snia{\hbox{{\rm SN~Ia}}}
\newcommand\snii{\hbox{{\rm SN~II}}}
\newcommand\fe{\hbox{{$Z_{\rm Fe}$}}}
\newcommand\solar{\hbox{{$Z_{\odot}$}}}
\newcommand\feml{\hbox{{$M_{\rm Fe}/L_{\rm B}$}}}
\newcommand\tx{\hbox{{$T_{\rm x}$}}}

\newcommand\bfrac{\hbox{{$f_{\rm b}$}}}
\newcommand\gfrac{\hbox{{$f_{\rm g}$}}}
\newcommand\omnot{\hbox{{$\Omega_{\rm m,0}$}}}
\newcommand\omb{\hbox{{$\Omega_{\rm b}$}}}
\newcommand\rvir{\hbox{{$r_{\rm vir}$}}}

\begin{document}

\articletitle{A Chandra and XMM View of the Mass \& Metals in Galaxy
Groups and Clusters}

\author{David A. Buote}

\affil{University of California, Irvine}
\email{buote@uci.edu}

\chaptitlerunninghead{Mass \& Metals in Galaxy
Groups and Clusters}

\begin{abstract}

X-ray observations with Chandra and XMM are providing valuable new
measurements of the baryonic and dark matter content of groups and
clusters. Masses of cD clusters obtained from X-ray and
gravitational lensing studies generally show good agreement,
therefore providing important validation of both methods. Gas
fractions have been obtained for several clusters that verify previous
results for a low matter density ($\omnot\approx 0.3$). Chandra has
also provided measurements of the mass profiles deep down into several
cluster cores and has generally found no significant deviations from
CDM predictions in contrast to the flat core density profiles inferred
from the rotation curves of low-surface brightness galaxies and dwarf
galaxies; i.e., there is no evidence for self-interacting dark matter
in cluster cores.  Finally, initial studies of the iron and silicon
abundances in centrally E-dominated groups show that they have
pronounced gradients from 1-2 solar values within the central 30-50
kpc that fall to values of 0.3-0.5 solar at larger radii. The Si/Fe
ratios are consistent with approximately 80\% of the metals
originating from Type Ia supernovae. Several cD clusters also display
central Fe enhancements suggestive of Type Ia supernova enrichment,
though some have central dips that may provide a vital clue for
solving the cooling flow mystery.

\end{abstract}


\section{Introduction}

It is now almost 4 years since the launches of the milestone X-ray
satellites, \chandra\ and \xmm. These missions possess greatly
improved capabilities over previous missions and, in particular, allow
for accurate spatially resolved spectral analysis of galaxy groups and
clusters. Initial results obtained for \chandra\ and \xmm\
observations of several low- and medium-redshift galaxy clusters and
some nearby groups of galaxies have now appeared in the literature. In
this review I will discuss some of these results, in particular as
they pertain to the mass distributions and metal abundances.

\section{Baryon Fractions \& Dark Matter in Clusters}

\subsection{Hydrostatic Equilibrium \& Lensing Comparisons}

The ability to measure the mass distributions of clusters from X-ray
observations requires that hydrostatic equilibrium be a suitable
approximation. Cosmological N-body simulations have been used to
address this issue and have determined that masses calculated by
assuming perfect hydrostatic equilibrium are generally quite accurate
provided the cluster in question is not currently experiencing a major
merging event (e.g., Tsai, Katz, \& Bertschinger 1994; Evrard et
al. 1996; Mathiesen et al. 1999). For individual clusters, the
assumption of hydrostatic equilibrium may be tested by comparing
masses obtained by X-ray and lensing methods, which serves as an
important check on both procedures.

Often the masses obtained from lensing studies (particularly strong
lensing) have exceeded the X-ray--determined values (e.g.,
Miralda-Escud\'{e} \& Babul 1995). For clusters that have regular
X-ray morphologies and usually house a cD galaxy wherein hydrostatic
equilibrium might be expected, it has been shown that the strong
lensing and X-ray methods can yield consistent masses provided the
X-ray emission is modeled by a multiphase spectrum characteristic of a
cooling flow (e.g., Allen 1998). Since, however, \chandra\ and \xmm\
have not found evidence for the cooling gas expected from standard
cooling flow models of clusters (Kaastra et al. 2001; Peterson et
al. 2001; Tamura et al. 2001), the interpretation of this agreement is
presently unclear.

The mass distributions of several clusters obtained from \chandra\
observations have been compared to determinations from strong and weak
lensing. For example, Allen, Schmidt, \& Fabian (2001) analyzed six
intermediate-redshift clusters and found good agreement in the masses
obtained from the (single-phase) X-ray and strong lensing methods.
Similarly good agreement between the weak-lensing and (single-phase)
X-ray analysis is also found by Arabadjis, Bautz, \& Garmire (2002)
for EMSS 1358+6245. In fact, weak-lensing determinations generally
agree with the X-ray values (e.g., A2218: Machacek et al. 2001; A1689:
Xue \& Wu 2002; CL0024+17: Ota et al. 2002).

In the cluster cores not all strong-lensing mass estimates agree with
the X-ray estimates from \chandra: A2218: Machecek et al. 2001, A1689:
Xue \& Wu 2002; CL0024+17: Ota et al. 2002. For these systems the
X-ray determined masses are a factor 2-3 lower than those determined
from the strong lensing. However, neither A2218 or CL0024+17 are cD
clusters but instead show substructure in the core both in the galaxy
and reconstructed mass distributions. A1689 does not display obvious
core substructure, though its strong arcs may suggest additional structure
aligned precisely behind the cluster core along the line-of-sight. 

Overall, the widespread agreement observed between the lensing and
X-ray methods outside of cluster cores is a strong validation of the
assumptions of both methods. The agreement of both methods in the
cores of relaxed cD clusters (except A1689) provides similar
validation of, in particular, the assumption of hydrostatic
equilibrium.  In the few cases of disagreement between strong lensing
and X-ray methods, substructure in the core is apparently responsible
-- though the case of A1689 remains to be resolved. Therefore, caution
must still be exercised when interpreting the results obtained from
either method, but the agreement between the methods for
morphologically regular clusters is quite encouraging.

\subsection{Gas Fractions and \omnot}

If the internal content of clusters is representative of the Universe
then, $\omnot = \omb/\bfrac$, where \omnot\ is the present matter
density, \omb\ is the present baryon density fixed by BBN calculations
and measurements of the deuterium abundance, and \bfrac\ is the
fraction of total baryons (gas+stars) in the cluster. Since it is
relatively easy to obtain an accurate measurement of the gas
contribution to the baryons using X-ray observations, this constraint
on the matter density is often written as $\omnot < \omb/\gfrac$,
where $\gfrac = M_{\rm gas} / M_{\rm cluster}$ is the total (hot) gas
fraction of the cluster; i.e., the gas fractions of clusters provide
an upper limit to \omnot.

Constraints on \omnot\ from baryon and gas fractions obtained from the
original studies of White et al. (1993) and White \& Fabian (1995)
were confirmed and refined with samples of 30-40 clusters using
\rosat\ data (e.g., Ettori \& Fabian 1999; Mohr, Mathiesen, \& Evrard
1999) and by SZ studies (e.g., Grego et al. 2001). The \rosat\ studies
took advantage of the improved imaging data with respect to previous
missions to obtain accurate gas density profiles. Unfortunately, the
temperature values crucial for calculating the gravitational mass
distributions had be taken from previously published emission-weighted
values obtained from broad-beam satellites. Nevertheless, precise
constraints were obtained and indicate a sub-critical matter density;
e.g., $\gfrac=0.212\pm 0.006$ and $\omnot < (0.36\pm 0.01)h_{50}^{-1}$
(Mohr et al. 1999).

Because accurate temperature profiles can now be obtained for clusters
with \chandra\ and \xmm, this analysis can be carried out with
the gas density, temperature, and therefore gravitating mass computed
consistently for each cluster from the same data set. Several
\chandra\ and \xmm\ studies have reported constraints on \omnot\ from
gas (and baryon) fractions. Allen et al. (2002) obtained
$\gfrac=0.113\pm 0.005$ within $r_{2500}$ in their sample of six
medium-redshift clusters for which masses obtained from the X-ray and
strong lensing methods agree to within 10\%. By making the assumption
that the baryon-to-hot gas ratio is the same as for Coma, they
inferred $\omnot = (0.32\pm 0.03)h_{70}^{-1}$ ($\Lambda$CDM).  Gas
fractions of 10\%-20\% are obtained from \chandra\ observations of,
e.g., Hydra-A (David et al. 2001), A1795 (Ettori et al. 2002), EMSS
1358+6245 (Arabadjis et al. 2002), and A2029 (Lewis, Buote, \& Stocke
2002). Similar results have been reported in \xmm\ studies: A1413
(Pratt \& Arnaud 2002), A1835 (Majerowicz, Neumann, \& Reiprich 2002),
and RX J1120.1 (Arnaud et al. 2002).

Overall, these gas fractions imply strong upper limits on the matter
density; e.g., $\omnot < 0.29\pm 0.03h_{70}^{-0.5}$ from the \chandra\
study of A2029 by Lewis et al. (2002).  The small inconsistencies in
the limits on \omnot\ obtained from gas fractions are due in part to
deviations from the assumed spherical symmetry and hydrostatic
equilibrium. Another contribution arises because different studies
quote gas fractions obtained at different radii. Since, however, it is
observed that gas fractions generally increase with radius, upper
limits quoted from gas fraction obtained at smaller radii should be
robust (provided hydrostatic equilibrium applies).

Hence, the results for gas fractions obtained from \chandra\ and \xmm\
indicate that $\omnot\approx 0.3$. This value is consistent with that
inferred from other methods (e.g., Tonry 2001; Freedman 2002).

\subsection{Dark Matter Properties}

There is presently considerable interest in the subject of the core
mass density profiles of galaxies and clusters because of what they
may tell us about the nature of the dark matter itself. Analyses of
the rotation curves of low surface brightness galaxies and dwarf
galaxies (e.g., Moore et al. 1999; Swaters et al. 2000) find that the
core mass density profiles ($\rho\propto r^{-\alpha}$) are generally
shallower ($\alpha\approx 0.5$) than the density profiles predicted
from CDM simulations of galaxy halos ($\alpha\approx 1-1.5$ --
Navarro, Frenk, \& White 1997 -- hereafter NFW; Moore et al. 1999).

If this discrepancy between observations and theory is real, it could
imply a fundamental problem with the CDM paradigm. Probably the most
widely investigated theoretical solution is the ``self-interacting''
dark matter model (SIDM) of Spergel \& Steinhardt (2000). In this
model the dark matter particles are assumed to posses some cross
section for elastic collisions with each other, and detailed CDM
simulations incorporating this idea have shown that indeed the dark
matter profiles of galaxies can be flattened as observed (e.g., Dav\'{e}
et al. 2001).

In principle galaxy clusters offer excellent venues to study dark
matter properties at a small fraction of a virial radius because they
are thought to be dominated by dark matter down to at least
$0.01\rvir$.  (e.g., Dubinski 1998). It is necessary to probe regions
deep into the core, $r=(0.001-0.01)\rvir$, to place the strongest
limits on any SIDM and to distinguish between competing
parameterizations of standard $\Lambda$CDM halos (e.g., Navarro 2001).
For massive clusters this corresponds to radii between a few kpc and a
few 10s of kpc. For nearby clusters ($z\la 0.2$) these scales are
resolved by \chandra, and thus it would seem the issue of the core
dark matter profiles can be addressed definitively with \chandra\
observations.

Unfortunately, nature has placed a roadblock that hinders
investigation of the core mass profiles of clusters. For clusters that
otherwise appear to be the most relaxed of their kind (i.e, those
usually referred to as ``cooling flows'' -- posses cD, X-ray images
regular outside $r=50-100$~kpc), the innermost core regions are
disturbed by the interaction of a central radio source with the hot
gas (e.g., McNamara 2002). These morphological disturbances certainly
raise doubts about hydrostatic equilibrium on the smallest scales and
therefore the ability of X-ray observations to address the core
density profile controversy.

Only a few papers have appeared in the literature on the subject of
the mass profiles of clusters determined from \chandra\ and \xmm\
observations. Most of these studies have not probed the inner cores
($r\la 0.01\rvir$) because the clusters are either too distant or the
(lower resolution) \xmm\ data were used. Generally, though, these
studies all find that CDM-like profiles (NFW and/or Moore) are
consistent with the data (Allen et al. 2002; Pratt \& Arnaud 2002;
Arabadjis et al. 2002).

In some clusters observed by \chandra\ to possess obviously disturbed
cores the mass profiles have been computed with varying results. David
et al. (2001) presented a detailed analysis of the core mass
distribution in Hydra-A which has a pronounced radio-X-ray disturbance
in the core. Nevertheless, they obtained a core density profile,
$\rho\sim r^{-1.3}$, in overall agreement with the NFW-Moore
parameterizations of CDM halos. In contrast, Ettori et al. (2002)
examined the core mass profile of A1795 which possesses filamentary
structure and an overall asymmetric X-ray core image. They obtained
$\rho\sim r^{-0.6}$ which is similar to the SIDM prediction and that
inferred for low-surface brightness and dwarf galaxies. However, these
authors have proposed a merger model which can account for the
morphological peculiarities while maintaining an NFW profile for the
gravitating matter of the cluster core.

Clearly, it is imperative to find clusters that are bright, nearby,
and undisturbed in their cores. Recently, Lewis et al. (2002) have
presented an analysis of the core mass distribution in the cD cluster
A2029 which meets these criteria. The \chandra\ image is highly
regular and displays none of the holes or other features observed in
most nearby otherwise-relaxed clusters. They find that the mass
distribution is well-constrained down to $\approx 0.003\rvir\approx
7$~kpc and is well-fit by the NFW profile. The mass profile does not
show any break within the half-light radius of the cD and is
consistent with the dark matter dominating over the entire region
probed. (Assuming a typical mass-to-light ratio for the stars, the
dark matter dominates down to $\approx 0.010\rvir$.) For this system,
therefore, the \chandra\ data clearly indicate no significant
contribution from SIDM. This suggests that the deviations from the CDM
predictions observed on small galaxy scales do not seem to imply a
fundamental problem with the general CDM paradigm. Instead, it is
likely that the numerical simulations do not currently account
properly for the effects of feedback processes on the formation and
evolution of small halos.

More \chandra\ observations of clusters with undisturbed cores are
very much needed.

\section{Metals}

\subsection{Groups}

\subsubsection{The Fe Problem}

There is presently a controversy regarding the iron abundances of
groups (and massive elliptical galaxies) deduced from X-ray
observations. While there seems to be general agreement of sub-solar
iron abundances outside the central regions ($r\approx 50-100$~kpc) of
groups (e.g., Finoguenov \& Ponman 1999; Buote 2000a), different
investigators have often obtained (for the same groups) different
results for the central regions ($r\la 50$~kpc). Most previous \rosat\
and \asca\ studies have found very sub-solar values of \fe\ in the
central regions of groups (for reviews see Buote 2000a; Mulchaey
2000). Since these low values of \fe\ are generally lower than the
stellar iron abundances (Trager et al. 2000), they imply that Type~Ia
supernovae (\snia) cannot have contributed significantly to the
enrichment of the hot gas. This implies that, unlike clusters, the
stellar initial mass function (IMF) in groups must be ``top heavy''
and very different from that of the Milky Way (e.g., Renzini et
al. 1993; Renzini 1997; Arimoto et al. 1997).  Consequently, various
authors have questioned the reliability of X-ray determinations of
\fe\ and suggested that the low \fe\ values are caused by errors
associated with the Fe L lines in X-ray plasma codes (e.g., Arimoto et
al. 1997; Renzini 2000).

However, in a series of papers (Buote \& Fabian 1998; Buote 2000a;
Buote 2000b) we found that indeed the iron abundances in the central
regions of groups were measured incorrectly, but not because of errors
in the plasma codes. Instead, we attributed the very sub-solar \fe\
values to an ``Fe Bias'' arising from forcing a single-temperature
model to fit a spectrum consisting of multiple temperature components
near 1 keV. The multiple temperature components may arise {\it either}
from the projection of single-phase gas from larger radii, or
represent real multiphase structure in the hot gas. Regardless of the
origin of the multitemperature structure, we find that $\fe/\solar\sim
0.7$ (or $\sim 1$ meteoritic solar) in the central regions of groups,
implying that a significant number of \snia\ have enriched the hot
gas, in better agreement with a Galactic IMF.

With the spatially resolved (medium resolution) spectroscopic
capabilities of the CCDs on \chandra\ and \xmm, as well as the
high-resolution spectral data of the centers afforded by the gratings
on both satellites, the Fe controversy can be finally settled. 

\subsubsection{Iron and Silicon from XMM Observations}

As of this writing we are aware of only two studies of the metal
abundances in galaxy groups with \xmm\ that are either published (NGC
1399: Buote 2002) or have been submitted for publication (NGC 5044:
Buote et al. 2002); and we are unaware of any \chandra\ publications
on this subject. The reason for the lack of \chandra\ publications is
probably due to the low-energy calibration problems that affect groups
with their $\sim 1$~keV temperatures more seriously than the higher
temperature clusters. Our initial analysis of the \xmm\ data of the
bright, nearby groups NGC 1399 and NGC 5044 have provided strong
constraints on the Fe and Si abundances as a function of radius: both
abundances take values between 1-2 solar within 30-50 kpc radii of
these systems and decrease to values 0.3-0.5 solar out to the largest
radii probed ($\sim 100$ kpc).

The super-solar central Fe values are obtained for both
single-temperature and two-temperature models, though the
two-temperature models are generally favored and give systematically
larger Fe abundances. The super-solar central Fe abundances obtained
with XMM also confirm that the very sub-solar values obtained in the
central regions of these and other groups with data from previous
satellites arose primarily from the Fe Bias caused by neglecting the
non-isothermal distribution of the hot gas; the previous underestimates
were also partially the result of using the wrong solar Fe abundance.
(We have also analyzed the \chandra\ data for NGC 5044 (Buote et
al. 2002) and find results for the Fe abundance that are overall
consistent with those obtained from the \xmm\ data. Although we
mention that the \chandra\ data for NGC 5044 tend to give Fe
abundances 10\%-30\% lower than those inferred from the \xmm\ data.)

These Fe and Si abundances measured with \xmm\ imply that
approximately 80\% of the Fe mass within r $\sim 50$~kpc originates
from Type Ia supernovae (SNIa). This SNIa fraction is similar to that
inferred for the Sun and therefore suggests a stellar initial mass
function similar to the Milky Way. Although this agreement with a
Galactic IMF may be satisfying to some, it should be emphasized that
detailed gas-dynamical models (without cooling flows) have difficulty
reproducing the observed radial dependence of Fe. Specifically, the
models (Buote et al. 2002) predict even larger Fe abundance values in
the central regions ($>3$ solar), a problem that dates back ten years
(e.g., Loewenstein \& Mathews 1991; Ciotti et al. 1991; Brighenti \&
Mathews 1999)

\subsection{Clusters}

Renzini and collaborators have shown that the iron mass-to-light ratio
(\feml) is approximately constant as a function of gas temperature
(\tx) for clusters but falls substantially for, $\tx\la 2$ keV,
corresponding to groups.  The constancy of \feml\ over the observed
range of \tx\ for clusters is consistent with the idea that star
formation has proceeded similarly in clusters and that Fe has not been
lost during their evolution (i.e., ``closed boxes'').  The latter
point is consistent with cosmological N-body simulations that find gas
fractions in clusters do not evolve appreciably (White et al. 1993;
Evrard 1997).

The global Fe abundances obtained from \chandra\ and \xmm\ studies of
clusters generally agree well with previous determinations from \asca\
and \sax. Many cD clusters show central enhancements of Fe (e.g., M87,
Hydra-A, A1795, A2029, A2199) suggesting \snia\ enrichment from the
cD. In fact, there are no reports of $\alpha$/Fe enhancements in
clusters that would imply enrichment dominated by \snii\ as was
claimed from early \asca\ studies (Mushotzky et al. 1996).

The effect of the Fe Bias in clusters has been demonstrated by Molendi
(2001) from analysis of the \xmm\ data for M87 which has the highest
data quality of any cluster. Molendi showed that the original results
reported for a radially varying single-temperature analysis
substantially underestimated the Fe abundance within the central
$30\arcsec$. After correcting for the bias the measured central Fe
abundance rose from $\approx 0.3\solar$ to $\approx 0.7\solar$, and
the previously reported central $\alpha$/Fe enhancements also
disappeared.

Finally, some clusters observed with \chandra\ have shown abundance
dips in the central regions. The best example is Centaurus for which
Sanders \& Fabian (2002) find that Fe rises to $\sim 1.8$ solar within
15~kpc of the center, then turns around and decreases back to $\sim
0.4$ solar in the central bin. This type of profile is not easily
understandable in terms of \snia\ enrichment from the central
galaxy. Morris \& Fabian (2002) have suggested the central dip is an
artifact of attempting to model a highly inhomogeneous metal
distribution with a homogeneous spectral model. This model is
attractive since it could explain why sub-keV cooling gas has not been
found by \chandra\ or \xmm\ in cooling flows.

\bigskip

\noindent I would like to thank the conference organizers both for 
inviting me to give this review and for arranging such an enjoyable
meeting.



\end{document}